\documentclass[sigconf]{acmart}

\usepackage[english]{babel}
\usepackage{fancyhdr}
\usepackage{xcolor}

\renewcommand\footnotetextcopyrightpermission[1]{} % removes footnote with conference info
\setcopyright{none}
\acmDOI{}

\acmISBN{}

\acmPrice{}

\begin{document}
\title{Linear Depth QFT over IBM Heavy-hex Architecture}

%\titlenote{Produces the permission block, and copyright information}
%\subtitle{Extended Abstract}

% \author{Paper \# XXX, XXX pages}
\author{Xiangyu Gao}
% \authornote{Note}
\affiliation{%
  \institution{New York University}
%   \streetaddress{Address}
%   \city{City} 
%   \state{State} 
%   \postcode{Zipcode}
}
\email{xg673@nyu.edu}
\author{Yuwei Jin}
% \authornote{Note}
\affiliation{%
  \institution{Rutgers University}
%   \streetaddress{Address}
%   \city{City} 
%   \state{State} 
%   \postcode{Zipcode}
}
\email{yj243@scarletmail.rutgers.edu}
\author{Minghao Guo}
% \authornote{Note}
\affiliation{%
  \institution{Rutgers University}
%   \streetaddress{Address}
%   \city{City} 
%   \state{State} 
%   \postcode{Zipcode}
}
\email{mg1998@scarletmail.rutgers.edu}
\author{Henry Chen}
% \authornote{Note}
\affiliation{%
  \institution{Rutgers University}
%   \streetaddress{Address}
%   \city{City} 
%   \state{State} 
%   \postcode{Zipcode}
}
\email{hc867@scarletmail.rutgers.edu}
\author{Eddy Z. Zhang}
% \authornote{Note}
\affiliation{%
  \institution{Rutgers University}
%   \streetaddress{Address}
%   \city{City} 
%   \state{State} 
%   \postcode{Zipcode}
}
\email{eddy.zhengzhang@gmail.com}

% \author{Firstname Lastname}
% \authornote{Note}
% \affiliation{%
%   \institution{Affiliation}
%   \streetaddress{Address}
%   \city{City} 
%   \state{State} 
%   \postcode{Zipcode}
% }
% \email{email@domain.com}

% \renewcommand{\shortauthors}{X.et al.}

\pagestyle{fancy}
\fancyhead{} % Clears all header fields

\begin{abstract}

Compiling a given quantum algorithm into a target hardware architecture is a challenging optimization problem. The compiler must take into consideration the coupling graph of physical qubits and the gate operation dependencies. The existing noise in hardware architectures requires the compilation to use as few running cycles as possible. Existing approaches include using SAT solver or heuristics to complete the mapping but these may cause the issue of either long compilation time (e.g., timeout after hours) or suboptimal compilation results in terms of running cycles (e.g., exponentially increasing number of total cycles).

In this paper, we propose an efficient mapping approach for Quantum Fourier Transformation (QFT) circuits over the existing IBM heavy-hex architecture. Such proposal first of all turns the architecture into a structure consisting of a straight line with dangling qubits, and then do the mapping over this generated structure recursively. The calculation shows that there is a linear depth upper bound for the time complexity of these structures and for a special case where there is 1 dangling qubit in every 5 qubits, the time complexity is 5N$+O(1)$. All these results are better than state of the art methods.

\end{abstract}

\maketitle
\section{Introduction}
\label{intro}

\begin{figure}[htb]
    \centering
    \includegraphics[width=0.49\textwidth]{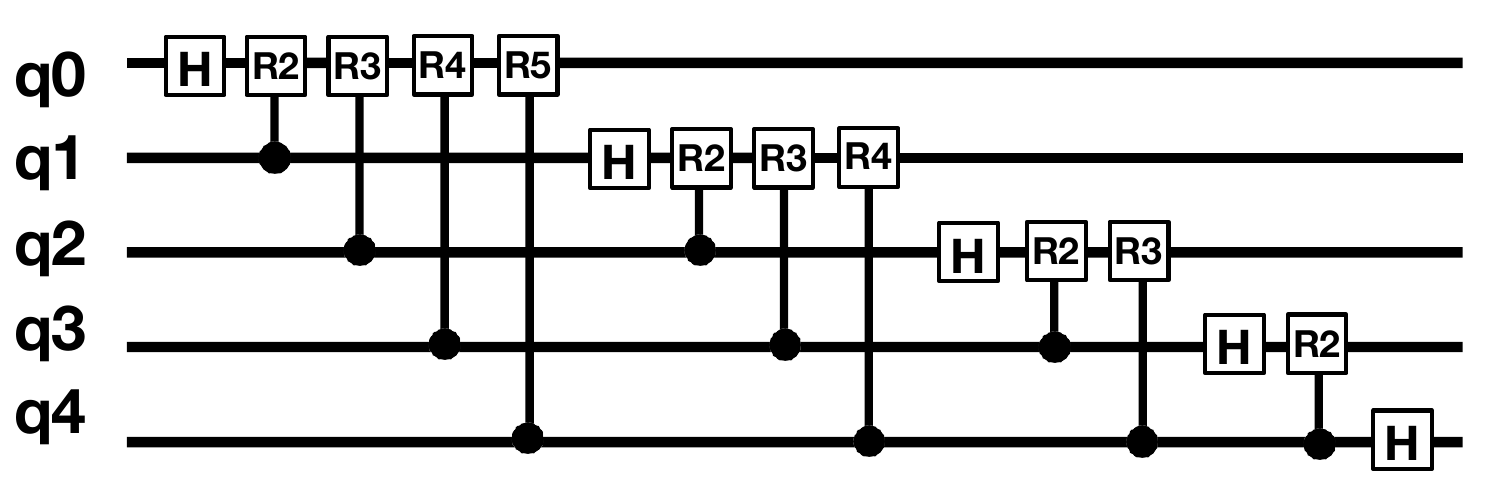}
    \caption{All necessary CPHASE gate operations over a QFT. H represents the Hadamard gate and Ri represents the CPHASE gate.}
    \label{fig:QFT}
\end{figure}

Quantum computing has become more and more popular across multiple fields, such as chemistry simulation, financial modeling, and cybersecurity. 
People develop quantum algorithms to realize many computations that cannot be completed within a reasonable amount of time using traditional computation methods.
One influential algorithm in this domain is Shor's algorithm~\cite{shor:focs94}, which brings people's attention to quantum computing because this algorithm can provide an exponential speed up to factor larger numbers into prime factors. 
This brings huge risks for several encryption methods, including the most popularly used RSA algorithm. Quantum Fourier Transformation (QFT) sits at the heart of Shor's algorithm, and therefore, it is one of the most important algorithms in quantum computing. In general, QFT requires all qubits to execute a Hadamard gate and all qubit pairs to execute a CPHASE gate. Figure~\ref{fig:QFT} shows a concrete example of all quantum gates for 5-qubit QFT.

The development of quantum computing also motivates a proliferation of quantum hardware architectures by industrial participants, including Google Sycamore, IBM heavy hex, LNN, and 2D grid. 
The concrete coupling graph varies from architecture to architecture. 
Typically, the number of connection links is relatively small with respect to the total number of qubits within the hardware compared with an all-to-all connection among all qubits. 
The compiler must consider the connection link as a constraint to do efficient circuit transformation and qubit mapping.

\textbf{Prior work} In this paper, we focus on the mapping algorithm for QFT over the IBM heavy-hex architecture. 
Prior work has tackled similar problems by developing general mapping algorithms for multiple architectures. 
However, they may cause the problem of long compilation time\cite{tan+:iccad20, tan+:iccad22, molavi+:micro22, murali+:asplos19}, or suboptimal mapping results~\cite{li+:asplos19}, or be too restricted to particular type of architectures with a Hamiltonian path to connect all qubits~\cite{maslov+:physreva07}. We cannot directly transplant those works to solve the QFT mapping over IBM heavy-hex architecture and generate satisfying outcomes.

\textbf{Our approach}. 
We generate a coupling graph of one major line with dangling qubits from heavy-hex architecture by deleting a tiny portion of connection links. 
Then, we label each qubit's number in increasing order from left to right, from up to bottom. 
The proposed QFT mapping algorithm leverages the proposed linear QFT by Maslov \cite{maslov+:physreva07} and Zhang \emph{et al} \cite{zhang+:asplos21} in linear nearest neighbor (LNN) architecture, with some modification. 
Based on our transformed heavy-hex coupling graph, we could implement the QFT mapping algorithm recursively. In our approach, we can have $m$ dangling nodes; each time there is a SWAP gate operation between a qubit in the major line and a dangling qubit, it reduces to the problem with $m - 1$ dangling nodes. The base case is the coupling graph with no dangling nodes, and this is exactly the same as the linear QFT scenario. 

\textbf{Contributions} Our contributions are multi-fold. We used a coupling graph tailored to IBM heavy-hex architecture containing repeated patterns. This coupling graph makes it possible for us to leverage a ``recursive" algorithm design to tackle the QFT mapping. Furthermore, we develop a mapping algorithm that takes linear complexity to map a QFT algorithm into the heavy-hex architecture. We have achieved $5N + O(1)$ complexity for a heavy hex architecture with $N$ qubits. Last but not least, we show the correctness of the proposal and calculate both the upper bound for a general case architecture and the exact time complexity for one special case. The head-to-head comparison offers significant improvement in the time complexity over an existing linear depth approach over the IBM heavy-hex architecture~\cite{jinxiangyu:24}.

\section{Qubit Mapping for IBM Heavy-hex}
In this section, we are going to describe the mapping algorithms for QFT over the IBM heavy-hex architecture in detail. 
We, first of all, present the concrete coupling graph of heavy-hex architecture;
then, based on the coupling graph, an initial mapping of logical qubits to physical qubits is provided to place each logical qubit in its initial position;
Finally, we provide a detailed algorithm to show the gate operations in each step during the QFT mapping process. 

\subsection{Coupling Graph}
\begin{figure}[htb]
    \centering
    \includegraphics[width=0.49\textwidth]{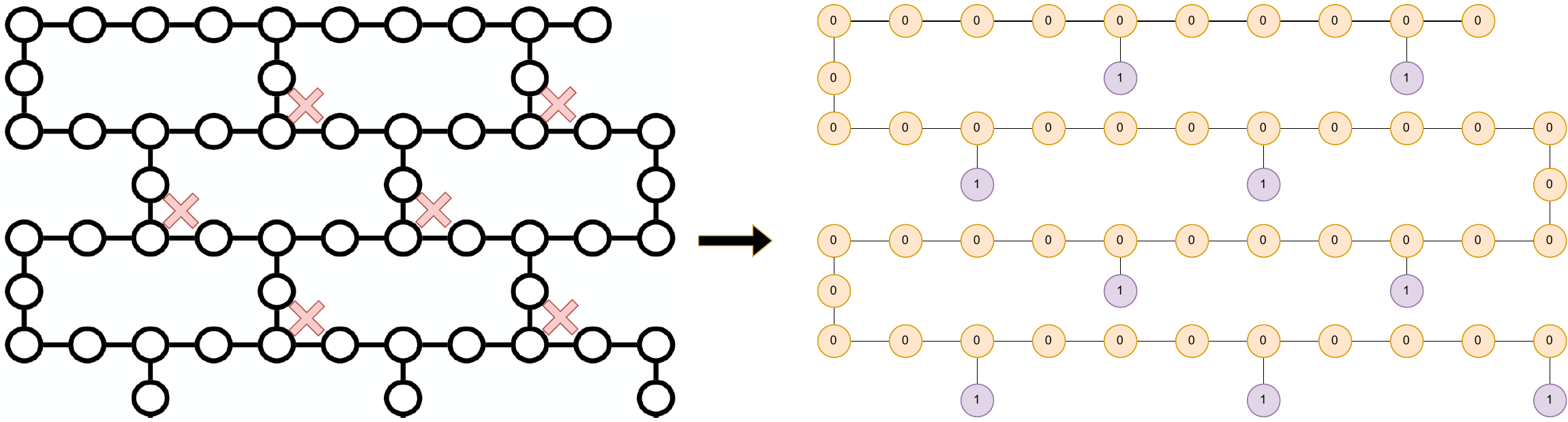}
    \caption{Turn heavy-hex (left-hand side) to our transformed coupling graph (right-hand side) with repeated patterns by removing some connection links (highlight by \textcolor{red}{X}). The transformed coupling graph consists of a major line (labeled by 0) with dangling points (labeled by 1).}
    \label{fig:IBM}
\end{figure}

\textbf{From heavy-hex to a generalized coupling graph.} 
We want the coupling graph to have repeated patterns so that the final solution can be generalized to an arbitrary number of such patterns with larger sizes. 
However, finding such patterns directly from the heavy-hex architecture is not straightforward.
% Our general approach to developing useful scheduling algorithms starts from a trial over a small size and then generalize it to larger sizes. 
% This means that building coupling graph with repeated patterns is a necessary step.
If we look at the original heavy-hex architecture on the left-hand side of Figure~\ref{fig:IBM}, it is not obvious to find repeated patterns directly from it. 
In other words, unlike LNN, 2D grid, or even Google Sycamore architecture where repeated patterns exist, we are unable to describe the given heavy-hex architecture by some parameter n.

To realize the goal, we simplify the coupling graph of the original heavy hex by deleting some connection lines in the architecture. 
Here is the current coupling graph. Each pattern consists of 2 parts: major line and dangling qubits. 
 All qubits with only one qubit as its neighbor will be defined as dangling qubits. All other qubits can form one straight line and are defined as the major line qubits.
As for a concrete heavy-hex example in Figure~\ref{fig:IBM}, the right-hand side shows a generated coupling graph with nine dangling nodes (labeled by 1), and all the rest (labeled by 0) are qubits in the major line.

Therefore, the architecture can be represented in a structured way using two types of qubits. This strikes a balance between qubit connectivity and pattern generalization. On the one side, it lets clear patterns exhibit from the heavy-hex architecture; on the other hand,  it keeps the majority of connection links, maximizing the utilization of the existing hardware architecture.

It is worthwhile to highlight some potential benefits of such coupling graph design, which would be leveraged in the mapping algorithm later. 
The coupling graph is close to LNN architecture. Most of all qubits are located in the major line, meaning they can be regarded as an extension of the LNN architecture. 
Besides, given a heavy-hex architecture, we have multiple ways to build the coupling graph by deciding which connection link to avoid. 
In the original setting, each ``dangling qubit" has two links to connect to 2 neighbor qubits. We can delete either of them to get the final transformed coupling graph.
This brings the possibility of error mitigation when particular links are prone to error.

\subsection{Definitions} 
We describe a few definitions before describing the initial mapping and the qubit mapping algorithm.

\begin{itemize}
    \item N1: the number of qubits on the major line.
    \item N2: the number of dangling qubits. 
    \item N: the total number of qubits. N $=$ N1 + N2.
    \item n: the total number of repeated groups/patterns in the coupling graph.
\end{itemize}

\subsection{Initial Mapping}

\begin{figure}[htb]
    \centering
    \includegraphics[width=0.49\textwidth]{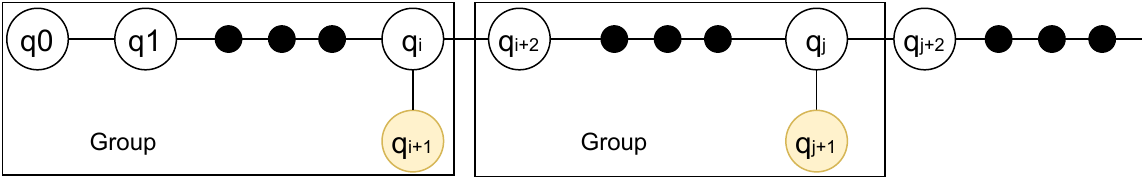}
    \caption{Initial mapping for Heavy-hex with dangling qubits: for any node $i$ that has both a node connected to it from the below and a node connected to it from the right (a ``T" junction node), the below node has index $i+1$, and the right node has index $i+2$. For any node $i$ that is not involved in a ``T" junction does not have a node below, its left node will be $i-1$, and the right node will be $i+1$.}
    \label{fig:initial_mapping_heavy_hex_dangling}
\end{figure}

We put logical qubits' initial mapping/placement in Figure~\ref{fig:initial_mapping_heavy_hex_dangling}.
In general, the qubits' number has an increasing order from left to right and up to down. 
The initial mapping should follow such a rule: when we compare any two adjacent qubits, the qubit in the right/down position has a larger index.   

We have also considered alternative mapping methods but they might cause potential problems. 
Suppose there are N qubits, N1 of which are placed in the major line, while the remaining N2 qubits are located as dangling points. 
One candidate choice is to label all qubits in the major line from 0 to N1 - 1 and the remaining qubits in the dangling positions from N1 to N - 1. This could lead to a problem where we have to relabel some of the qubits if we increase the number of groups in our particular design. 
In our case, the existing initial mapping method guarantees that adding a new group (a set of nodes with one dangling point) will not affect the previous qubit mapping. This brings potential benefits to calculating the time complexity by induction since we could do this with a focus on the extra steps required from the added group) rather than doing everything from scratch every time.

\subsection{Qubit Mapping Algorithm}

Before describing the algorithm details, we define a few terms in this domain. We can see from Figure~\ref{fig:recursive} that there are two types of nodes in our defined coupling graph of heavy-hex. The first type contains $N1$ qubits, and we call it the \textbf{major line}, while the second type contains the remaining $N2$ qubits dangling from the major line. So we call them \textbf{dangling points}. 

\subsubsection{Revisit the linear QFT mapping algorithm}
\begin{figure}[htb]
    \centering
    \includegraphics[width=0.49\textwidth]{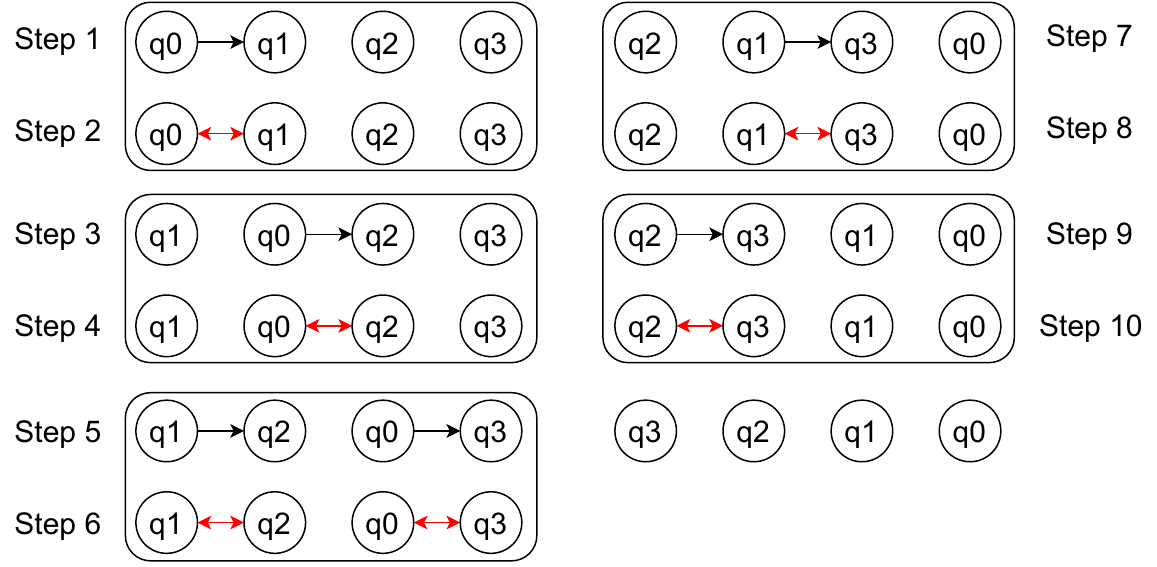}
    \caption{The concrete steps in QFT mapping for LNN over a line with 4 qubits. Each pair consists of one CPHASE step and SWAP step occurring in the same connection links.}
    \label{fig:LNN}
\end{figure}

Before describing our proposal for QFT mapping over the heavy-hex architecture, we would revisit the QFT mapping over linear nearest neighbor(LNN) structure~\cite{zhang+:asplos21} because there would be several similarities. 
In general, QFT mapping over LNN has interleaving steps between CPHASE cycle and SWAP cycle. For each pair consisting of CPHASE and SWAP, they would appear in the same link in two consecutive steps. 
A concrete example with 4 qubits in a line is shown in Figure~\ref{fig:LNN}. We could see that the final mapping after QFT ($q_3$ to $q_0$) is a mirror image of the initial mapping ($q_0$ to $q_3$).

\subsubsection{Special Case of Only One Dangling Point}

Let's consider the special case when there is only one dangling node from the major line. We will extend it to the case of multiple dangling points later. 

Below are concrete steps of our QFT mapping solution. We start running QFT as if it is on a line of nodes from $q_{0}$ to $q_{i+1}$ first until a certain point.

\begin{figure}[htb]
    \centering
    \includegraphics[width=0.49\textwidth]{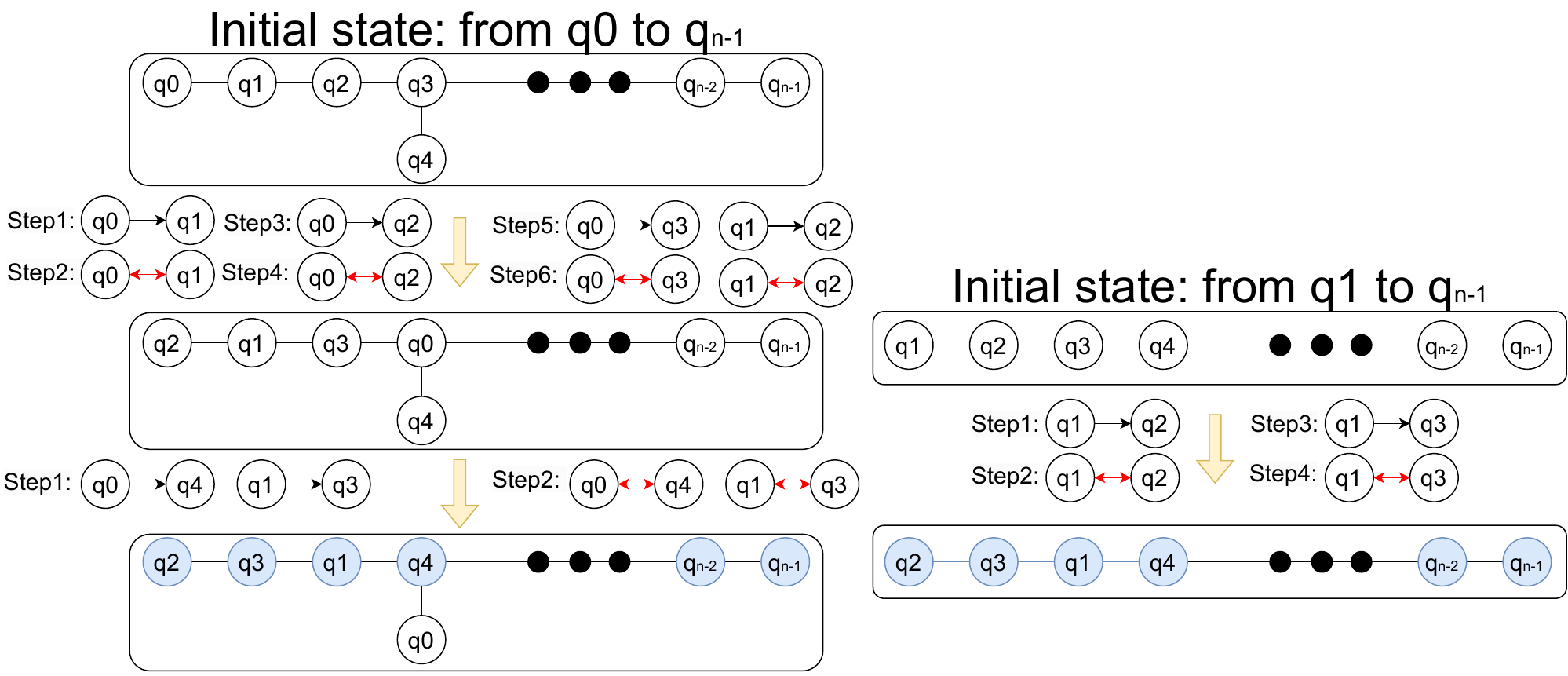}
    \caption{Rules of gate operations for the QFT mapping with one dangling point. The LHS shows the process of turning the QFT mapping into 3 phases: before $q_0$ moves to the place above $q_4$, interaction between $q_0$ and $q_4$, all steps afterwards. The initial state of the third phase is an intermediate step of RHS where we do QFT for $q_1$ to $q_{n-1}$, realizing the reduction from one dangling point to none.
    % \eddy{Expand the caption.}
    }
    \label{fig:swap_then_control_heavy_hex}
\end{figure}

\begin{figure}[htb]
    \centering
    \includegraphics[width=0.49\textwidth]{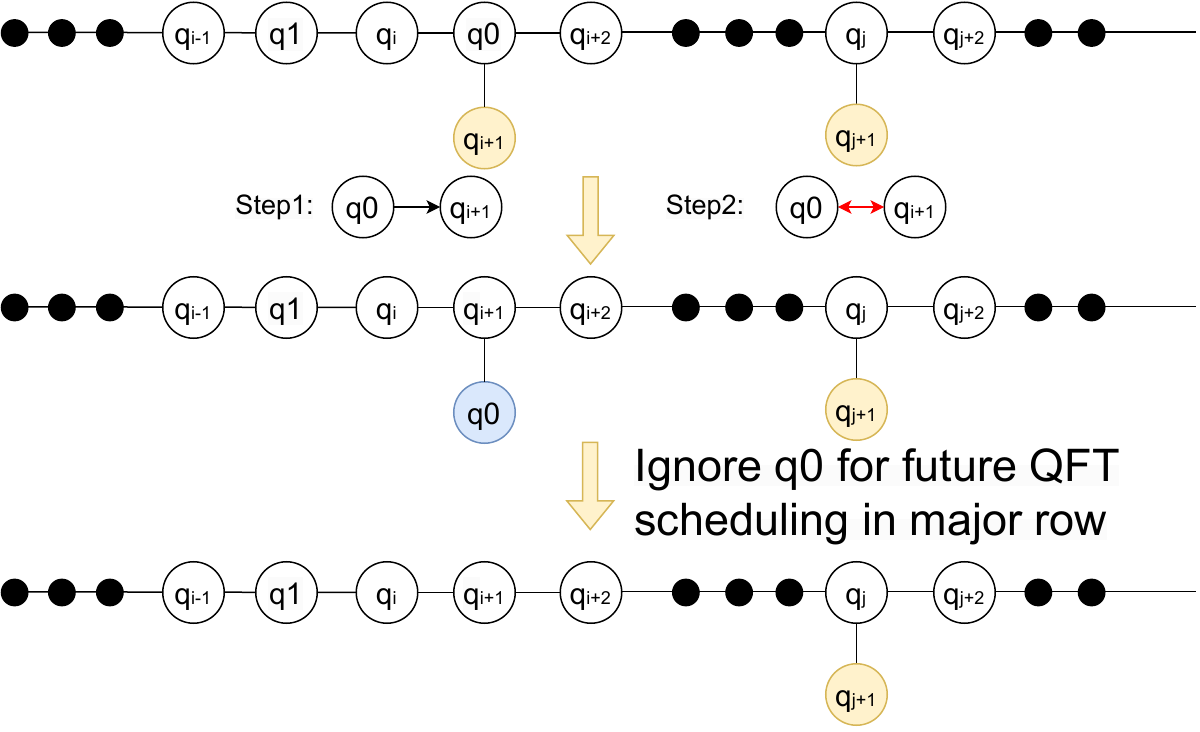}
    \caption{Recursive pattern: reducing one dangling node from the coupling graph after two steps: one CPHASE gate and one SWAP gate between the qubit above and the dangling point itself.}
    \label{fig:recursive}
\end{figure}

\begin{itemize}

  \item The first time when q0 is above (connected to) $q_{i+1}$ (the dangling point), as shown in Figure \ref{fig:recursive}, we will perform SWAP(q0, $q_{i+1}$). Then, the positions of qubits of the major line are similar to that in an intermediate step when we do linear QFT over a straight line starting from q1 to $q_{N1 - 1}$, and the timing is right before $q_1$ swaps with $q_{i+1}$. 
  \item Then let it continue with the steps of doing QFT for $q_{1}$ to $q_{N1-1}$ on a straight line until done. Note that whenever a qubit $q_j$ such that $ j > i $ is moved to next to $q_0$ in the dangling position for the first time, we pause and let $q_0$ perform a CPHASE with it. And then the algorithm is done. 

\end{itemize}

A concrete example is shown in Figure \ref{fig:swap_then_control_heavy_hex}, where the dangling point $i+1$  is set as 4.

From the above, one can see what all nodes from $q_1$ to $q_{N1-1}$ can perform CPHASE gates using the usual linear QFT solution. What about the node $q0$? Since we moved it to the dangling position, we kept it there until the end of the algorithm. Before it is moved to the dangling position, it has interacted with qubits $q_1$ to $q_{i}$. Then how do we ensure it also performs CPHASE with nodes from $q_{i+1}$ to $q_{N1-1}$? In the straight line QFT, as shown in prior work \cite{zhang+:asplos21, maslov+:physreva07}, each qubit $i$ moves towards the left first, stops for one step, and then to the right, for $N1-1$ distance all  altogether (we define the distance between two adjacent qubits as 1), where $N1$ is the length of the major straight line. All nodes after $q_{i+1}$ on the original line are to the right of $q_0$'s fixed location in the dangling point. And these nodes haven't started moving when $q_0$ is moved to the dangling location. Each of them needs to move $N1-1$ distance. Among all such nodes, the farthest node to $q_0$'s dangling location has a distance of $N1-i-1$, smaller than $N1-1$.  Hence when these nodes move to the left, they will be above $q_0$ at one time point. Whenever they meet $q_0$, we let $q_0$ perform a CPHASE with these qubits. Hence the interaction between $q_0$ and the nodes from $q_{i+2}$ to $q_{N1-1}$ are completed.

Now the problem is solved; all qubits interact. The dependence is also satisfied. We sketch the proof here. As $q_0$ interacts with these nodes from $q_{i+2}$ to $q_{N1-1}$, due to the fact that a node moves faster than a node that is originally on its right. However, $q_1$ could interact with $q_{i+2}$ to $q_{N1-1}$ earlier. This is okay. It is because the true dependence in QFT, is that a gate ($q_i$, $q_j$) where $q_i$ is the control and $q_j$ is the target, has to execute after ($q_k$, $q_i$) $k<=i$. Since ($q_0$, $q_{i+2}$) and ($q_1$, $q_{i+2}$) do not have dependence, they can run in any order. Moreover, the time ($q_1$, $q_k$) where $k>1$ runs, ($q_0$, $q_1$) has already happened, as $q_1$ has already moved past $q_0$.

\subsubsection{Extending to the Case of More than One Dangling Points}

 Now we describe the case of more than one dangling point. We show how it is handled in a recursive manner. Specifically, we could reduce the number of dangling nodes by one each time. Let's assume we have two nodes dangling from the major line, as shown in Figure \ref{fig:recursive}, $q_{i+1}$ and $q_{j+1}$. We follow the solution that is the same as the single-dangling-point case. The first time we swap $q_0$ with $q_{i+1}$, it is as if we can remove $q_0$ from the entire line, and then it reduces to the problem of having only one dangling point. We continue the solution for a straight line with one dangling point, and the problem is solved.  Note that $q_0$ is in a fixed position now (the first dangling position), and all the remaining nodes to the right of $q_{i+1}$ will move to the left first, and they can interact with $q_0$.

 Hence, we proved the case from one to two dangling points. The same idea works for from $k$ dangling points to $k+1$ dangling points. The reason is that the previous $k$ qubits $q_0$ to $q_{k-1}$ have already moved to the fixed dangling locations (as if the are already removed), and the remaining sequence on the major line starts from $q_{k}$, and that is reduced to the single-dangling-node problem again. And we can prove that $q_{k-1}$, whenever it is moved to the $k$-th dangling location, it will later interact with all nodes to the right of it in the original placement.

We haven't talked about CPHASE gates yet. For the straight-line QFT case, the CPHASE happens in the same way as the linear QFT. 
A special scenario is when a right qubit (right of the original locations of the first $k$ qubits) reaches the ``T-junction" for the first time, where there is a dangling qubit below. If dependence is satisfied, we let the qubit above and below in the ``T-junction" perform a CPHASE gate. 

\subsection{Special Case}

Figure~\ref{fig:swap_then_control_heavy_hex} is a special case where each group consists of 5 qubits such that we define a group of 5 qubits as that 4 of which are in the major line and one is in the dangling position. This is similar to what we have unrolled from the heavy-hex architecture. There is a dangling qubit for every four (or eight) qubits on the major line. We will show the complexity result of that later. 

\section{Case Study and General Complexity}
\begin{figure}[htb]
    \centering
    \includegraphics[width=0.49\textwidth]{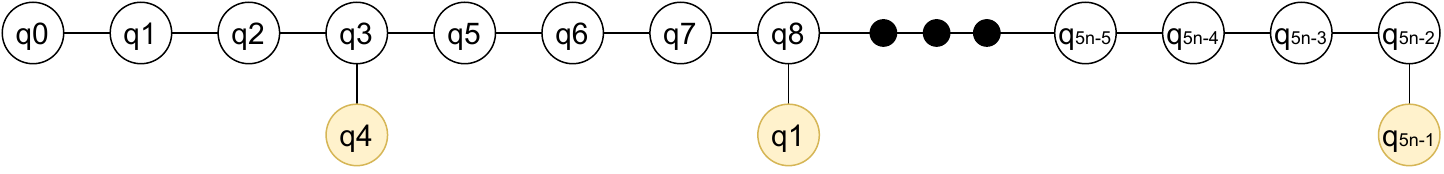}
    \caption{Initial mapping for Heavy-hex with 5n qubits (with 5 qubits per group).}
    \label{fig:initial_mapping_heavy_hex}
\end{figure}

\begin{figure}[htb]
    \centering
    \includegraphics[width=0.49\textwidth]{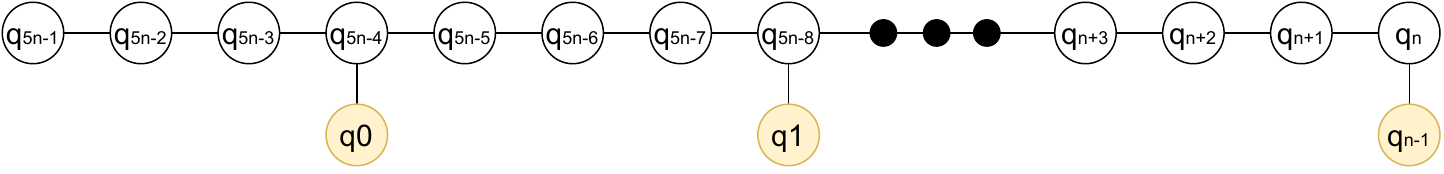}
    \caption{Final mapping for Heavy-hex with 5n qubits (with 5 qubits per group).}
    \label{fig:final_mapping_heavy_hex}
\end{figure}

In this section, we will first present a detailed analysis of a special case where there will be a dangling qubit in every five qubits with initial mapping in Figure~\ref{fig:initial_mapping_heavy_hex} and final mapping in Figure~\ref{fig:final_mapping_heavy_hex}. 
Correctness analysis and time complexity analysis are given to show that our proposed algorithm can correctly complete with the time complexity to be $5N+O(1)$ where N is the total number of qubits within the hardware architecture. Later, we will show that for general cases where there is an arbitrary number of dangling points, the complexity is at most $6N+O(1)$.

\subsection{Correctness Analysis}
Guaranteeing the correctness of our implementation is always the first priority.
We verify the correctness of QFT gate execution in 2 major aspects. First, we check whether or not all CPHASE gates have been executed. Specifically, in a heavy-hex architecture with N qubits, there should be one CPHASE gate per qubit pair and hence unduplicated $\frac{N(N-1)}{2}$ CPHASE gates in total. 
In addition, all these gate executions should follow the dependency constraints in QFT, e.g., any $q_{i}$ cannot take the control qubit role before the completion of all CPHASE gates where $q_{i}$ is the target qubit.

\textbf{Gate count verification.} As is shown in Table~\ref{table:simulation}, the simulator collects the information of $\#$ depth, $\#$ SWAP and the $\#$ CPHASE. It passes the verification that $\#$ CPHASE = $\frac{N(N-1)}{2}$ empirically. 

\begin{figure}[htb]
    \centering
    \includegraphics[width=0.49\textwidth]{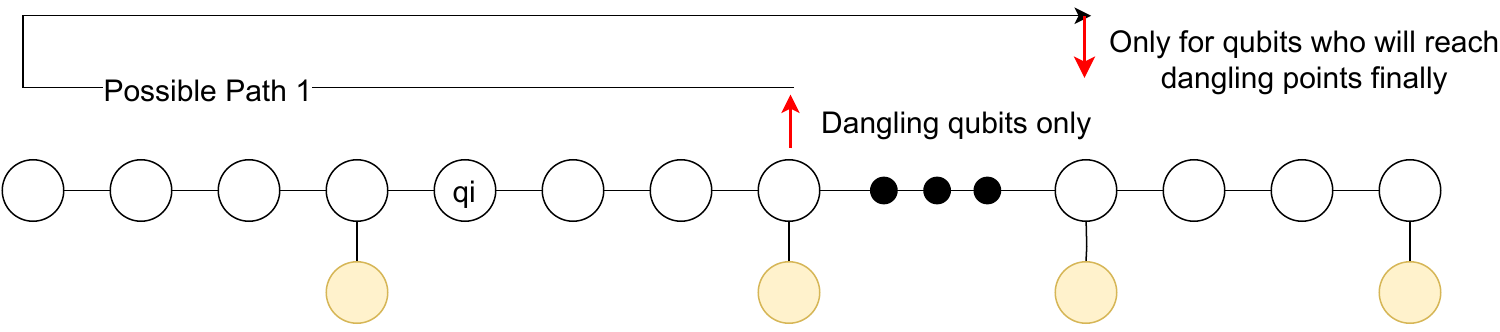}
    \caption{All possible paths for each qubit during QFT mapping. Black path is required for all qubits while \textcolor{red}{red path} is only for qubits whose initial/final position is in dangling nodes.}
    \label{fig:qubit_path}
\end{figure}

\textbf{Gate operation dependency verification.} 
Furthermore, we should guarantee no duplicate CPHASE operations and that all gate operations follow the dependency relationship. 
As we can check from our proposal, 
The proof starts from understanding the path of each qubit.
Figure~\ref{fig:qubit_path} plots all possible paths for each qubit during the QFT mapping process.
In general, each qubit has at least two phases: moving to the left until it hits the end and then moving in the right direction. Two additional actions may happen, such that one node moves down to a dangling point or out of a dangling point. 

When one qubit moves to the left, it will meet all qubits whose number is smaller than itself; every time it is adjacent to another qubit with a smaller index, there is a CPHASE gate execution in between. Therefore, before reaching the leftmost position, all qubits have completed the CPHASE gate operations, where they act as target qubits. As this qubit moves to the right, it would play the control-qubit role in CPHASE gate operation. This ensures all necessary CPHASE gate operations follow the gate operation dependency described before and are executed once and exactly once during our scheduling algorithm.

In terms of those qubits whose initial or final positions are in the dangling position. Depending on the initial place of the qubits and the qubit's index, it might need an extra step to move up from the dangling position and an extra step to move down to the dangling position (its final destination). This would not cause any issues either because each time one qubit arrives at a position where the qubit below has a larger number, there would be a CPHASE gate followed by a SWAP gate in between, which sets the qubit with a smaller number to its final destination and initiates the move for the qubit with a larger number. Then it will follow the mapping order described in the paragraph above.

Such analysis proves the correctness of our proposed QFT mapping algorithm. 

\begin{table}[h!]
\begin{tabular}{|l|l|lll|}
\hline
\multicolumn{1}{|c|}{Architecture} & \multicolumn{1}{c|}{\# qubits} & \multicolumn{3}{c|}{Our approach} \\ \hline
          &     & \multicolumn{1}{l|}{Depth} & \multicolumn{1}{l|}{\# SWAP} & \# CPHASE \\ \hline
Heavy-hex & 2*5 & \multicolumn{1}{l|}{39}    & \multicolumn{1}{l|}{40}      & 45       \\ \hline
Heavy-hex & 3*5 & \multicolumn{1}{l|}{64}    & \multicolumn{1}{l|}{90}      & 105      \\ \hline
Heavy-hex & 4*5 & \multicolumn{1}{l|}{89}    & \multicolumn{1}{l|}{160}     & 190      \\ \hline
Heavy-hex & 5*5 & \multicolumn{1}{l|}{114}   & \multicolumn{1}{l|}{250}     & 300      \\ \hline
Heavy-hex & 6*5 & \multicolumn{1}{l|}{139}   & \multicolumn{1}{l|}{360}     & 435      \\ \hline
Heavy-hex & 7*5 & \multicolumn{1}{l|}{164}   & \multicolumn{1}{l|}{490}     & 595     \\ \hline
Heavy-hex & 8*5 & \multicolumn{1}{l|}{189}   & \multicolumn{1}{l|}{640}     & 780     \\ \hline
\end{tabular}
\begin{small}
\caption{Resource usage of our approach over heavy-hex configurations with different \# qubits. Assuming there are 5 qubits per group, 4 of which are in major line and the remaining one is the dangling point. 
% \textbf{TODO: in the future, Compare against SATMAP and SABRE}
\label{table:simulation}} 
\end{small}
\end{table}

\subsection{Time Complexity}

\subsubsection{Complexity for this special case}
\begin{figure}[htb]
    \centering
    \includegraphics[width=0.49\textwidth]{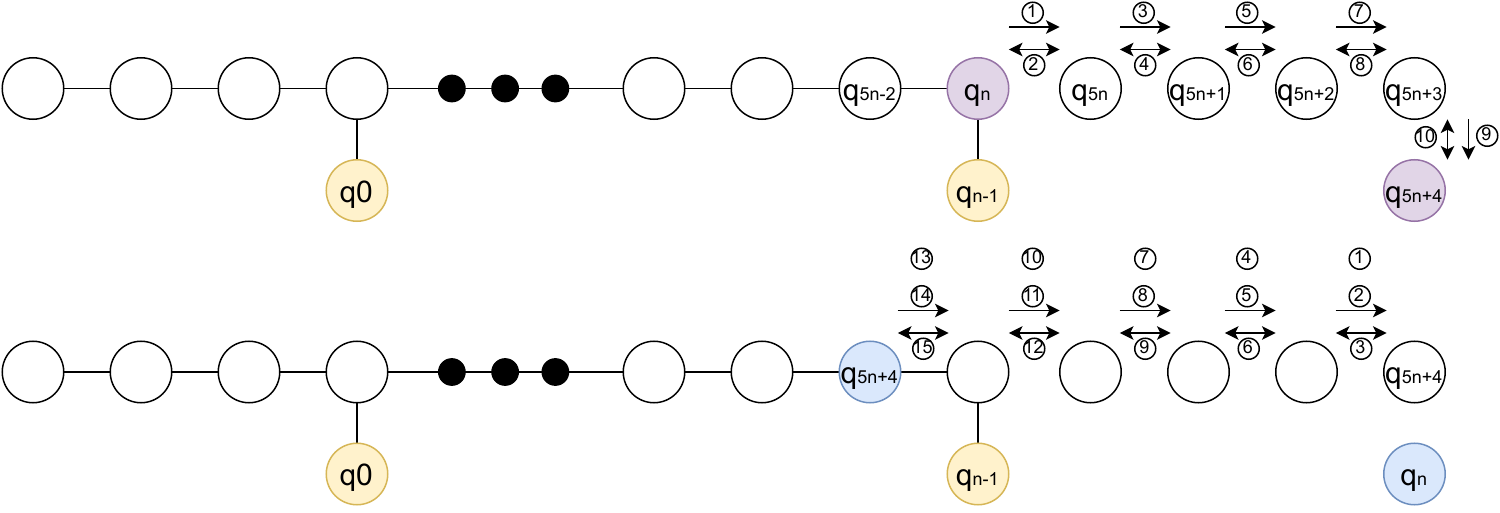}
    \vspace{-0.1cm}
    \caption{Extra steps required for one added qubit group. The top figure shows the steps in the first phase, while the bottom figure shows that in the second phase.}
    \label{fig:extra_steps}
\end{figure}

One benefit of our initial mapping design is that when adding a new qubit group into the architecture, the labeling of the previous qubits keeps unchanged. This means that a lot of gate operations will be overlapped. Specifically, comparing heavy-hex architecture with 5n qubits vs that with 5(n+1) qubits, several steps would execute the same or similar gate operations. For instance, in Figure~\ref{fig:extra_steps}, in a heavy-hex architecture with 5(n+1) qubits, before $q_{n-1}$ arrives at its final location (the penultimate member of the dangling positions), every gate operation in each step identical with that architecture with 5n qubits.

\begin{figure*}[htp]
    \centering
    \includegraphics[width=0.86\textwidth]{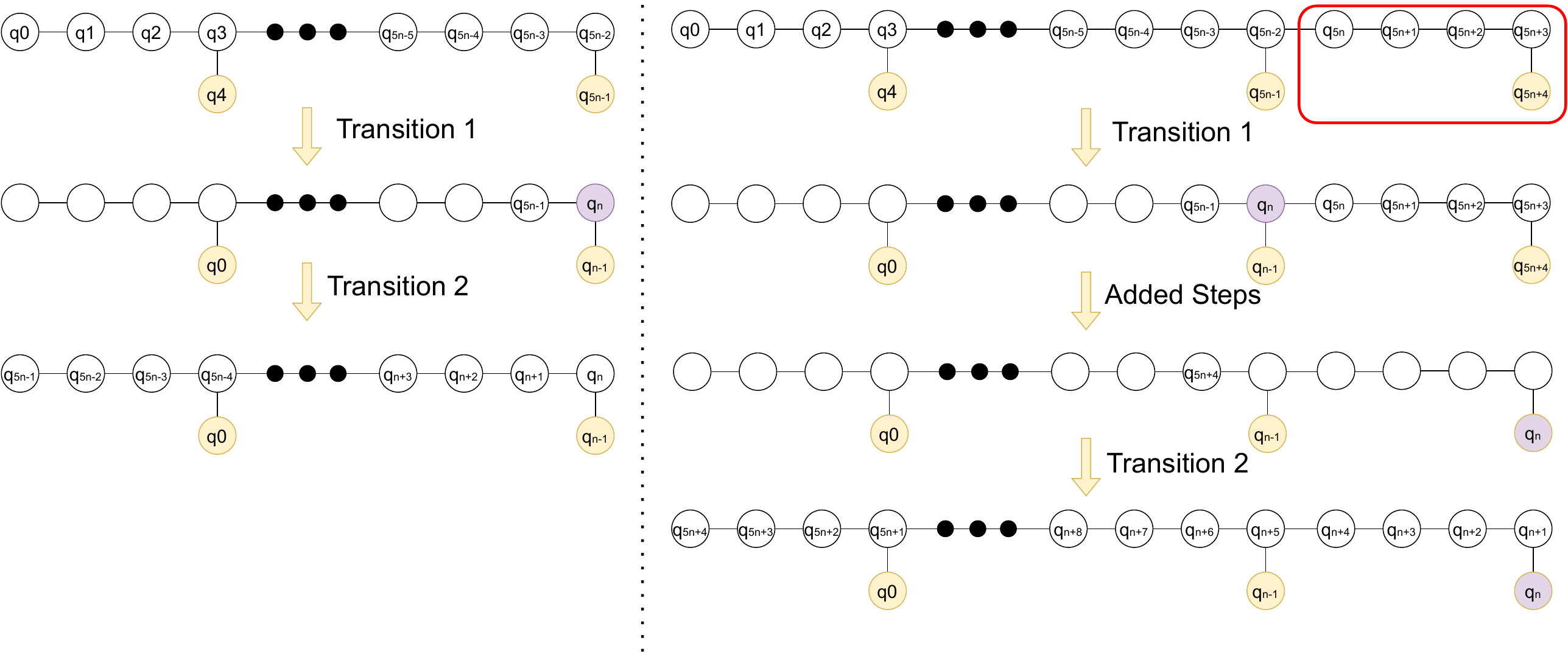}
    \vspace{-0.3cm}
    \caption{The place where extra steps appear. The LHS is the QFT mapping for architecture with n groups while the RHS is for architecture with (n+1) groups. They share the same cycles in transition 1 and transition 2 but the RHS has to use added steps to complete additional gate operations for newly added qubits.}
    \label{fig:extra_steps_for_new_group}
    \vspace{-0.3cm}
\end{figure*}

Hence, instead of directly calculating the time complexity, we could leverage the architecture features in our initial mapping and only focus on \textbf{the extra steps} required for an added qubit group. Then, the time complexity could be achieved by induction proof. 

Figure~\ref{fig:extra_steps_for_new_group} decouples the QFT mapping into multiple states and captures the added portion if we add one more group into the architecture. To get the extra steps, we should pay attention to just the operation over the \textbf{physical qubits} rather than the logical qubits. The LHS of Figure~\ref{fig:extra_steps_for_new_group} divides the QFT mapping for 5n qubits into three states: initial state, the state when the qubit with the biggest number $q_{5n-1}$ is one step before $q_{n}$, and the final state. 
Let's use $Q_{k}$ to label the physical qubit where $q_{5n-1}$ sits after transition 1.
Compared with the version where there are 5n + 5 qubits in the RHS, the added steps occur to move $q_{n}$ to its final destination after transition 1 and move the qubit with the biggest number $q_{5n+4}$ to physical qubit $Q_{k}$. The remaining steps are overlapped.

As for the added steps, we divide them into 2 phases shown in Figure~\ref{fig:extra_steps}. The first phase is to move $q_{n}$ forward until its final destination is located at the end of dangling positions; the second phase is to move $q_{5n+4}$ backward to the places where $q_{5n-2}$ sits before phase 1. There are ten steps in the first phase to move $q_{n}$ forward until the last location of the dangling positions, and another 15 steps are necessary in the second phase to move $q_{5n+4}$ backward to the correct place. 

After summing up the steps of these 2 phases, here is the conclusion that each time we add another pattern that consists of 5 qubits, we need an extra 25 steps to complete the QFT mapping. Therefore, as the number of qubits grows, the time complexity would be 5N $+O(1)$, where N is the total number of qubits in the heavy-hex architecture.

Considering the fact that the QFT mapping algorithm in~\cite{zhang+:asplos21} with complexity 4N+O(1) works for LNN architecture where we could connect all qubits within one line, the target of our architecture is a much more complex coupling graph so the proposed QFT approach completes with a slightly higher time complexity 5N+O(1).

\subsubsection{Upper bound for general cases.} In addition to the 
We also calculate the upper bound for a general case where there are N1 qubits in the major line with N2 dangling nodes (label them as $q_{k_{0}}$ ... $q_{k_{N2-1}}$) in the dangling positions. However, the distance between 2 nearby dangling nodes might be different.

In order to calculate the upper bound, we could decompose the whole QFT mapping process into multiple parts. There could be potential opportunities to merge steps among different parts, but the focus here is to get the upper bounds. We leave the next-step optimization to future work. The concrete description of the different parts is as follows:

Part I: QFT mapping in the major line only. 
As for this part, we could leverage the results from~\cite{zhang+:asplos21}. It shows that the QFT mapping over an LNN consists of phases with only CPHASE gate operations and with SWAP operations. 
In a straight line with N1 qubits, there are 2*N1-3 CPHASE cycles and 2*N1-3 SWAP cycles. Therefore, the total number of cycles is 4*N1-6.

Part II: interaction between the major line and dangling points. 
This part can be divided into two categories. 
The first category is related to the steps required to SWAP the qubit in the major line with the dangling point above. Specifically, given there are N2 dangling points, the final mapping shows that qubits from $q_0$ to $q_{N2-1}$ should finally be located in the dangling position. This means a SWAP operation is required to change the location between $q_{i}$ and the dangling qubit above when it reaches the ``T junction" above its final destination from the left. In addition to the SWAP operation, there is a CPHASE operation before, so for each of these N2 qubits, it needs an extra 2 phases to do such interaction, and the total time complexity is 2*N2;
the second category is related to the CPHASE gate operation between dangling points and major lines where the dangling qubit is used as the control qubit. These steps should be added when the major line is doing LNN QFT mapping. As we can see from the existing QFT mapping over LNN, there will be interleaving between CPHASE and SWAP cycles. The CPHASE cycle will do CPHASE gate operations, not changing the relative qubit position while SWAP cycle will change some qubits' position. Therefore, the extreme case is that every time we have a SWAP cycle, we need to stop for another cycle to allow this second category of CPHASE gate operations. The total time complexity is 2*N1-3.

By summing up the complexity from Part I and II, we could get the upper bound of our approach, which is 6N1 - 9 + 2N2. Since there is a known fact that $N1 + N2 = N$, $N1 \leq N$, and $N2 \leq N$, the upper bound time complexity is 2N+4N1+O(1) $\leq$ 6N+O(1). As for a sanity check, in our special case where N1=$\frac{4}{5}$N, we could get the upper bound time complexity to be 5.2N+O(1), which is bigger than the actual time complexity, 5N+O(1).

\section{Conclusion}

We introduce a new approach to do QFT scheduling over the currently emerging IBM heavy-hex architecture by considering it as a coupling graph with major row and dangling qubits.
Compared with state-of-the-art approaches, the scheduling result we generate uses fewer usage of circuit depth.
We hope our results will encourage 
engineers to transplant such an idea to do qubit mapping for more emerging quantum algorithms over this architecture in the future.

\bibliographystyle{ACM-Reference-Format}
\bibliography{reference}

\end{document}